\documentclass[12pt]{article}

\usepackage{amsmath,amssymb}
\usepackage{graphicx}

\makeatletter

\renewcommand\section{\@startsection{section}{1}{\z@}
                                   {-3.5ex \@plus -1ex \@minus -.2ex}
                                   {2.3ex \@plus .2ex}
                                   {\normalfont\large\bfseries}}
\renewcommand\subsection{\@startsection{subsection}{2}{\z@}
                                   {-3.25ex\@plus -1ex \@minus -.2ex}
                                   {1.5ex \@plus .2ex}
                                   {\normalfont\normalsize\bfseries}}
\renewcommand\subsubsection{\@startsection{subsubsection}{3}{\z@}
                                   {-3.25ex\@plus -1ex \@minus -.2ex}
                                   {1.5ex \@plus .2ex}
                                   {\normalfont\normalsize\bfseries}}
\renewcommand\paragraph{\@startsection{paragraph}{4}{\z@}
                                   {3.25ex \@plus1ex \@minus.2ex}
                                   {-1em}
                                   {\normalfont\normalsize\bfseries}}

\makeatother

\newdimen\tableauside\tableauside=1.0ex
\newdimen\tableaurule\tableaurule=0.4pt
\newdimen\tableaustep
\def\phantomhrule#1{\hbox{\vbox to0pt{\hrule height\tableaurule
width#1\vss}}}
\def\phantomvrule#1{\vbox{\hbox to0pt{\vrule width\tableaurule
height#1\hss}}}
\def\sqr{\vbox{%
  \phantomhrule\tableaustep

\hbox{\phantomvrule\tableaustep\kern\tableaustep\phantomvrule\tableaustep}%
  \hbox{\vbox{\phantomhrule\tableauside}\kern-\tableaurule}}}
\def\squares#1{\hbox{\count0=#1\noindent\loop\sqr
  \advance\count0 by-1 \ifnum\count0>0\repeat}}
\def\tableau#1{\vcenter{\offinterlineskip
  \tableaustep=\tableauside\advance\tableaustep by-\tableaurule
  \kern\normallineskip\hbox
    {\kern\normallineskip\vbox
      {\gettableau#1 0 }%
     \kern\normallineskip\kern\tableaurule}%
  \kern\normallineskip\kern\tableaurule}}
\def\gettableau#1 {\ifnum#1=0\let\next=\null\else
  \squares{#1}\let\next=\gettableau\fi\next}

\newcommand{\be}{\begin{equation}}
\newcommand{\ee}{\end{equation}}
\newcommand{\bea}{\begin{eqnarray}}
\newcommand{\eea}{\end{eqnarray}}
\newcommand{\ba}{\begin{array}}
\newcommand{\ea}{\end{array}}

\newcommand{\id}{\hbox{1\kern-.27em l}}
\newcommand{\ZZ}{\mathbb{Z}}
\newcommand{\RR}{\mathbb{R}}
\newcommand{\half}{ {\textstyle \frac{1}{2}  } }

\newcommand{\al}{\alpha}
\newcommand{\ga}{\gamma}

\newcommand{\bet}{\beta}

\newcommand{\ka}{\kappa}
\newcommand{\de}{\delta}

\newcommand{\si}{\sigma}
\newcommand{\la}{\lambda}

\newcommand{\ze}{\zeta}

\newcommand{\tha}{\theta}

\newcommand{\Ups}{\Upsilon}

\newcommand{\cN}{\mathcal{N}}

\newcommand{\D}{{\rm d}}

\newcommand{\rar}{\rightarrow}
\newcommand{\emp}{\emptyset}
\newcommand{\non}{\nonumber}

\newcommand{\SU}{\mathrm{SU}}
\newcommand{\SO}{\mathrm{SO}}

\newcommand{\Sp}{\mathrm{Sp}}
\newcommand{\su}{\mathrm{su}}
\newcommand{\so}{\mathrm{so}}
\newcommand{\spl}{\mathrm{sp}}

\newcommand{\Spin}{\mathrm{Spin}}
\newcommand{\Pin}{\mathrm{Pin}}

\begin{document}

\begin{center}

\vspace*{5mm}
{\Large\sf  Rigid surface operators and S-duality: some proposals}

\vspace*{5mm}
{\large Niclas Wyllard}

\vspace*{5mm}
Department of Fundamental Physics\\
Chalmers University of Technology\\
S-412 96 G\"oteborg, Sweden\\[3mm]
{\tt wyllard@chalmers.se}          

\vspace*{5mm}{\bf Abstract:} 

\end{center}

\noindent 
We study surface operators in the $\cN=4$ supersymmetric Yang-Mills theories with gauge groups $\SO(n)$ and  $\Sp(2n)$. As recently shown by Gukov and Witten these theories have a class of rigid surface operators which are expected to be related by S-duality. The rigid surface operators are of two types, unipotent and semisimple. We make explicit proposals for how the S-duality map should act on unipotent surface operators. We also discuss semisimple surface operators and make some proposals for certain subclasses of such operators.

\setcounter{equation}{0}
\section{Introduction}

Surface operators in gauge theories are natural generalisations of the Wilson and 't~Hooft operators (which are based on curves/lines). Surface operators were almost completely overlooked for a long time. Part of the reason was  that there were no clear applications of such operators as compared to the more well-known  Wilson and 't Hooft operators.  Recently Gukov and Witten initiated a study of surface operators~\cite{Gukov:2006} (see also \cite{Witten:2007} for a short review and references). Although the discussion in \cite{Gukov:2006} is carried out for a specific gauge theory ($\cN=4$ super-Yang-Mills) with a specific application in mind, surface operators are expected to be a generic feature in gauge theories. 

The $\cN=4$ supersymmetric Yang-Mills theories may well be the simplest (gauge) quantum field theories in $3+1$ dimensions. These theories have a large number of symmetries and special features. One such symmetry is the mysterious S-duality symmetry. 

The $S$-duality conjecture \cite{Montonen:1977} for the  $\cN = 4$ supersymmetric four-dimensional Yang-Mills theories states that the theory with gauge group $G$ and a value of the complexified coupling constant 
$
\tau = \frac{\theta}{2 \pi} + \frac{i}{g_{\rm YM}^2},
$
where $\theta$ is the theta angle and $g_{\rm YM}$ is the Yang-Mills coupling constant, is equivalent to the theories arising from the transformations $S$ and $T$: 
\bea \label{S}
S : \; (G, \tau) & \rightarrow & ( G^{\vee}, - 1 / r \tau)\, , \non \\
T : \; (G, \tau) & \rightarrow & (G, \tau + 1) \,,
\eea
where $G^{\vee}$ denotes the dual group of $G$ \cite{Goddard:1976} and $r$ is the ratio of the lengths-squared of the long and short roots of the Lie algebra of $G$ (see e.g.~\cite{Argyres:2006} for a recent discussion). For the simple groups with simply-laced Lie algebras, $G^{\vee}$ and $G$ are equal at the Lie algebra level. However, this is not true for all groups. 
Some examples of  S-dual pairs, that will be studied further in this paper, are:
\be
\begin{array}{llcl}
\underline{G} & \underline{G^{\vee}} && \underline{C} \vspace*{2mm} \cr
\Spin(2n{+}1) & \Sp(2n)/\ZZ_2 && \ZZ_2 \cr
\Sp(2n) & \Spin(2n{+}1)/\ZZ_2\equiv \SO(2n{+}1)  && \ZZ_2 \cr
\SO(2n) & \SO(2n) && \ZZ_2 \,.
\end{array}
\ee
Here $C$ denotes the centre of the group $G$. 

The S-duality conjecture is well established, but has not been proven, and it is in general difficult to devise tests of the conjecture. One common strategy is to look for objects that are independent of the coupling constant and hence should have a counterpart in the dual gauge theory. 

In a recent paper \cite{Gukov:2008} Gukov and Witten extended their earlier analysis of surface operators and identified a subclass of surface operators in the $\cN=4$ super-Yang-Mills theories which preserve half the supersymmetries and have the property that they are {\it rigid} (which essentially means that they can not be changed by an adiabatic change of $\tau$). Rigid surface operators therefore provide a class of operators that are expected to be closed (i.e.~related to each other) under S-duality. (S-duality properties of other classes of surface operators have been studied in \cite{Gukov:2006,Gomis:2007}.)

It was shown in \cite{Gukov:2008} that the rigid surface operators are of two types: unipotent and semisimple. The rigid semisimple surface operators in the theories with gauge groups $\SO(n)$ and  $\Sp(2n)$   are labelled by pairs of certain partitions. Unipotent rigid surface operators arise in the limit when one of the two partitions is empty. 

Partitions have also appeared in other recent works on 
S-duality~\cite{Henningson:2007a,Wyllard:2007}. These works have in common that they count quantum states. For such states one can have quantum-mechanical state mixing which complicates the search for an S-duality map. Therefore in~\cite{Henningson:2007a,Wyllard:2007} only the total number of states with certain quantum numbers were counted.  The rigid surface operators on the other hand  appear not to suffer from such quantum ambiguities and it therefore makes sense to look for an S-duality map, mapping a rigid surface operator in the theory with gauge group $G$ into a rigid surface operator in the theory with gauge group $G^\vee$. 
In \cite{Gukov:2008} the search for such an S-duality map was begun and some proposals for the S-duality map relating rigid surface operators in the $B_n$ ($\SO(2n{+}1)$) and $C_n$ ($\Sp(2n)$) theories for low ranks were made. A certain special subclass of unipotent rigid surface operators was also argued to be closed under S-duality.  
In addition,  a problematic mismatch in the total number of rigid surface operators in the $B_n$ and $C_n$ theories was pointed out. 

In this paper we attempt to extend the analysis begun in \cite{Gukov:2008}. 
In particular, we make several proposals for how the S-duality map should act on certain classes of rigid surface operators in the $\cN=4$ $B_n$ and $C_n$ theories.
We also make some comments and proposals for the $D_n$ ($\SO(2n)$) theories. 

In the next section we review the construction of rigid surface operators given in \cite{Gukov:2008} and discuss some mathematical results and definitions that will be needed in later sections. We also discuss certain invariants of the surface operators, i.e.~expressions that are expected to be unchanged under the S-duality map. In particular, we review the invariants proposed in \cite{Gukov:2008} and also propose a new invariant, which is closely related to `fingerprint' invariant discussed in \cite{Gukov:2008}.  
Then in section \ref{BC} we discuss rigid surface operators in the $B_n$ and $C_n$ theories and make several proposals for how the S-duality map should act on certain classes of surface operators. In particular, we make a proposal for how the S-duality map should act on  unipotent rigid surface operators. We also discuss semisimple surface operators and the  mismatch of the total number of rigid surface operators and try to find a way to characterise the problematic surface operators.
Finally, in section \ref{D} we briefly discuss the $D_n$ theories and make a proposal for how the S-duality map should act on unipotent rigid surface operators and also discuss a class of semisimple surface operators.
In the appendix we tabulate, as an example, all rigid surface operators and their associated invariants in the $\SO(13)$ and $\Sp(12)$ theories.

\setcounter{equation}{0}
\section{Surface operators in $\cN=4$ super-Yang-Mills}\label{surf}

The $\cN=4$ super-Yang-Mills theory is a four-dimensional gauge theory with gauge group $G$ and the following field content: a gauge field (1-form), $A_\mu$ ($\mu=0,1,2,3$), four Majorana spinors $\psi^a$ ($a=1,2,3,4$) and six real scalars, $\phi_I$ ($I=1,\ldots,6$). All fields take values in the adjoint representation of the gauge group. 

Surface operators are generalisations of the Wilson and 't Hooft operators in gauge theories. Instead of being localised on a one-dimensional submanifold they are localised on a two-dimensional surface. The definition of  surface operators in \cite{Gukov:2006,Gukov:2008} involves a generalisation of the definition of 't Hooft operators (see also \cite{Drukker:2008} and references therein for a discussion of various ways to define surface operators).

A surface operator is defined by prescribing a certain singularity structure of the gauge (and scalar) fields near the surface on which the operator is supported. We only consider surface operators supported on a $\RR^2$ submanifold (denoted $D$) of flat four-dimensional space. The surface $D$ is taken to lie at $x_2=x_3=0$ and the gauge 1-form in the directions normal to the surface is $A=A_2\, \D x^2 +A_3\, \D x^3$.  
To preserve half of the supersymmetries, the full $\SO(6)$ R symmetry group can not be unbroken. By selecting two of the six scalars in the $\cN=4$ super-Yang-Mills theory ($\phi_2$ and $\phi_3$ say) and forming $\phi = \phi_2 \,\D x^2 + \phi_3 \,\D x^3$, the conditions for preserving half of the supersymmetries can be written \cite{Gukov:2008}  
\bea \label{hitch}
&&F - \phi \wedge \phi = 0 \,, \non \\
&&\D\phi + A \wedge \phi + \phi \wedge A = 0\,, \\
&& \D \star\! \phi + A \wedge \star \phi + \star \phi \wedge A = 0 \, , \non 
\eea
where $F =\D A + A\wedge A$ as usual. The equations (\ref{hitch}) are known as Hitchin's equations. A solution to these equations with a prescribed singularity along the surface $D$ defines a surface operator.

Up to gauge transformations the most general rotation-invariant Ansatz for $A$ and $\phi$ is (here $x_2+ix_3 = re^{i\tha}$)
\bea
A &=& a(r) \, \D \tha \,, \non \\
\phi &=& c(r) \, \D \tha + b(r) \frac{\D r }{r}  \,, \\
\star\phi &=& -b(r) \, \D \tha  + c(r) \frac{\D r }{r} \non \,.
\eea
Inserting this Ansatz into (\ref{hitch}) and defining $s = \ln r$ one finds that (\ref{hitch}) reduce to Nahm's equations:
\bea \label{nahm}
\frac{\D a}{\D s} = [b,c]\,, \non \\
\frac{\D b}{\D s} = [c,a] \,,\\
\frac{\D c}{\D s} = [a,b] \,. \non
\eea
If one is interested in conformally invariant surface operators one naively expects that scale invariance would require that $a,b,c$ have to be independent of $s$ ($r$). Nahm's equations then imply that the constant elements $a$, $b$ and $c$ need to mutually commute. Surface operators of this type were treated in \cite{Gukov:2006}. The new insight in \cite{Gukov:2008} was to point out another way to obtain conformally invariant surface  operators. 

Nahm's equations (\ref{nahm}) are solved by 
\be \label{nahmsol}
a = \frac{T_x}{s + 1/f}\,,\qquad b = \frac{T_z}{s + 1/f}\,,\qquad c = \frac{T_y}{s + 1/f} \,,
\ee
provided that 
\be \label{su2}
[T_x, T_y] =  T_z \qquad \mbox{et cycl.} 
\ee
i.e.~the $T_i$'s span a representation (in general reducible) of the  $\su(2)$ Lie algebra. The  $T_i$'s also have to belong to the adjoint representation of the gauge group. 

It would seem that the surface operator obtained from the solution (\ref{nahmsol}) depends on $f$ (for a fixed $f$). However, in \cite{Gukov:2008} it was argued that one should think of $f$ as being allowed to fluctuate. Then provided certain additional constraints (to be discussed below) are fulfilled, the resulting surface operator does not depend on any parameters and therefore has to be scale invariant. It is expected that it is in fact also superconformal. 

Another way to characterise the surface operators can be obtained by considering the conjugacy class (orbit under gauge conjugation) of the monodromy
\be
U = P \exp(\oint \mathcal{A}) \,,
\ee
where $\mathcal{A} = A + i \phi$ and the integration is around a circle with constant $r$, near $r=0$. Note that $U$ belongs to the complexified gauge group and the conjugacy class is therefore a conjugacy class in the complexified gauge group. Note also that $\mathcal{F} = \D \mathcal{A} + \mathcal{A}\wedge \mathcal{A}=0$, which follows from (\ref{hitch}) and means that $U$ is unchanged under deformations of the integration contour. For the solution (\ref{nahmsol}) $U$ becomes 
\be \label{Uplus}
U= P \exp(\frac{2\pi}{s+1/f} \,\,  T_+ ) \,,
\ee
where $T_+\equiv T_x +i T_y$ is nilpotent (strictly upper (or lower) triangular in matrix language). A conjugacy class of this type is called unipotent (the corresponding Lie algebra orbit is called nilpotent).  

The above construction of surface operators does not exhaust all possibilities \cite{Gukov:2008}. This can be seen by noting that there are two types of conjugacy classes in a Lie group: unipotent and semisimple. Above we only discussed unipotent classes. However, semisimple classes can also lead to rigid surface operators. 
The above discussion can be modified to incorporate  semisimple conjugacy classes using the following construction. Consider a semisimple (diagonalisable in matrix language) element $S$ of the gauge group and require that near the surface $D$, 
\be \label{Scond}
S \Ups(r,\tha) S^{-1} = \Ups(r,\tha+2\pi) \,,
\ee 
for all adjoint-valued fields $\Ups$ in the theory. This effectively breaks the gauge group to the centraliser of $S$ (i.e.~all group elements which commute with $S$). One can combine this with the above construction by looking for a solution to Nahm's equations which in addition also satisfies, near $r=0$, the restriction arising from $S$, (\ref{Scond}). At the level of conjugacy classes this combination of the two constructions means that one considers more general monodromies of the form $V=SU$, where $S$ is semisimple and $U$ is unipotent.

From the above discussion we see that what is needed to find the possible surface operators is a classification of unipotent and semisimple conjugacy classes. In general the construction of surface operators from conjugacy classes leads to a large variety of surface operators not all of which are expected not to depend on any parameters and to have a simple behaviour under S-duality. What is needed is a criteria which can be used to decide  when a surface operator is `rigid'.  

Nilpotent orbits (unipotent conjugacy classes) have been classified by mathematicians.  A nilpotent/unipotent orbit whose dimension is strictly smaller than that of any nearby orbit is called rigid. All rigid orbits have been classified (see \cite{Gukov:2008} and chapter 7 of \cite{Collingwood:1993} for further details). This result will be reviewed for the classical groups in the next subsection.

There exist semisimple conjugacy classes which have the property that the centraliser (unbroken gauge group) of such a class is larger than that of any nearby class (such classes are called isolated in the mathematics literature, see e.g.~\cite{Hezard:2004}, chapter 2).
The possible isolated classes $S$ were obtained in~\cite{Gukov:2008} (see also section 4.1.2 in \cite{Hezard:2004}); for the classical groups, this result will be reviewed in the next subsection.

Surface operators based on monodromies of the form $V=SU$, where $S$ is semisimple and isolated and $U$ is unipotent and rigid will be called  rigid and are expected to be superconformal and not to depend on any parameters and to have a simple behaviour under S-duality. The classification of rigid surface operators in the theories with  classical gauge groups will be discussed in the next subsection. 

In \cite{Gukov:2008} a distinction is made between strongly rigid and weakly rigid surface operators. Throughout this paper we will only consider strongly rigid operators which we for simplicity simply refer to as rigid surface operators. The larger class including also the weakly rigid surface operators could possibly be useful in resolving some of the unsolved problems.

\subsection{Some mathematical definitions and results}
We saw above that rigid surface operators correspond to certain (unipotent and semisimple) conjugacy classes of the (complexified) gauge group. We summarise below the main mathematical results and definitions that will be needed in this paper. A readable mathematics reference is \cite{Collingwood:1993}. We will describe in  detail the rigid surface operators in the theories with classical gauge groups. Since the $A_n$ series does not have any non-trivial rigid surface operators we will concentrate on the $B_n$, $C_n$ and $D_n$ series. 

It is always possible to choose a block-diagonal basis for $T_+$ (cf.~(\ref{Uplus})),
 \be 
 \label{Ti}
 T_+ = \left( \ba{ccc} T_+^{n_1}  & & \\ 
                        & \ddots &   \\
                        & & T_+^{n_l}
 \ea \right ),  
 \ee 
where $T_+^{n_k}$ is the `raising' generator of the $n_k$-dimensional
irreducible representation of $\su(2)$. 
For the $A_n$ series (i.e.~$\SU(n{+}1)$ gauge groups) the above argument gives the complete solution, but for 
 the other classical groups, i.e.~$B_n$ ($\SO(2n{+}1)$), $C_n$ ($\Sp(2n)$) and $D_n$ ($\SO(2n)$), there are restrictions on the allowed dimensions of the $\su(2)$ irreps arising from the requirement that $T_+$ should belong to the relevant gauge group. 
This problem has been solved by mathematicians; see also \cite{Naculich:2001} for a discussion in the Physics literature (the authors of this publication were unaware of the fact that the problem had been solved by mathematicians decades earlier). The unbroken gauge Lie algebra (i.e.~the subalgebra commuting with the $\su(2)$ generators) has also been worked out. 

For $\SO(n)$ ($\Sp(2n)$) $\sum_{k=1}^{l} n_k$ equals $n$ ($2n$) and 
the restrictions on the building blocks ($\su(2)$ irreps) and unbroken Lie algebra are summarised in the following table:

\begin{center}  
\begin{tabular}{|c|c|c|} 
\hline
Gauge group & Allowed $\su(2)$ representations & gauge enhancement \\ 
\hline
 & &  \\[-13pt]
\hline 
 & &  \\[-13pt]
$\Sp(2n)$   & $2m$ odd-dimensional irreps 
          & $\spl(2m)$ \\[1pt]
\cline{2-3}
 & &  \\[-13pt]
{} & $m$ even-dimensional irreps & $\so(m)$  \\
 & &  \\[-13pt]
\hline
& &  \\[-13pt]
\hline
& &  \\[-13pt]
$\SO(n)$  & $2m$ even-dimensional irreps 
           & $\spl(2m)$  \\[1pt]
\cline{2-3}
& &  \\[-13pt]
{} & $m$ odd-dimensional irreps & $\so(m)$ \\ 
\hline
\end{tabular}  
\end{center} 

From the block-decomposition (\ref{Ti}) we see see that unipotent (nilpotent) surface operators are classified by partitions. The fact that not all $\su(2)$ representations are allowed means that the classification involves restricted partitions.

A partition $\la$ of the positive integer $n$ is a collection of positive integers, $\la_i$, (the parts of the partition) such that $\sum_{i=1}^l \la_i = n$. We use the convention that $\la_1\ge \la_2 \ge \cdots \ge \la_l$. The integer $l$ (the number of parts of the partition) is called the length of the partition. Throughout this paper we use a short-hand notation to denote partitions. For instance $3^32^41$ corresponds to $3+3+3+2+2+2+2+1$. Partitions can be added in an obvious way. If $\la$ and $\ka$ are partitions then $\la+\ka$ is the partition with parts $\la_i+\ka_i$. 
Partitions are in a one-to-one correspondence with Young tableaux. For instance the partition $3^32^41$  corresponds to
\be
\tableau{3 7 8} 
\ee
An  {\it orthogonal} partition is a partition where all even integers appear an even number of times. A {\it symplectic} partition is a partition for which all odd integers appear an even number of times. 
An orthogonal (symplectic) partition is called {\it rigid} if it has no gaps (i.e.~$\la_i-\la_{i+1}\leq1$ for all $i$) and no odd (even) integer appears exactly twice. Rigid unipotent surface operators in the $B_n$ and $D_n$ theories are in one-to-one correspondence with rigid orthogonal partitions of $2n{+}1$ and $2n$, respectively.  Rigid unipotent surface operators in the $C_n$ theories are in one-to-one correspondence with rigid symplectic partitions of $2n$. (See \cite{Gukov:2008} for more details.)

The transpose of a partition is the partition obtained by interchanging the roles of the rows and columns of the Young tableau. For instance
\be
\left(\tableau{1 3 4} \right)^t \quad = \quad  \tableau{1 2 2 3} 
\ee
The transposed partition is again a partition, but if the original partition belongs to some restricted class of partitions then the transposed partition may or may not belong to the same class. 

In the theories under consideration, a partition $\la$ is called {\it special} if the following condition holds
\bea
B_n: &\quad \la^t &\;\mbox{is orthogonal} \,,\non \\
C_n: &\quad \la^t &\;\mbox{is symplectic} \,, \\
D_n: &\quad \la^t &\;\mbox{is symplectic} \,.\non 
\eea
In particular, these definitions imply that for the $B_n$ case all rows in the Young tableau corresponding to a rigid special partition have to be odd, whereas for the $C_n$ and $D_n$ cases all rows in the Young tableau corresponding to a rigid special partition have to be even. 

A partition is called {\it rather odd} if any odd integer appears at most once.

For the $B_n$ , $C_n$ and $D_n$ theories it has been proven \cite{Gukov:2008,Hezard:2004} that the possible isolated semisimple conjugacy classes (cf.~discussion above) correspond to diagonal matrices, $S$, with the only allowed elements along the diagonal being $+1$ and $-1$. The possible  matrices $S$ break the gauge group in the following way (at the Lie algebra level)
\bea
\so(2n{+}1) &\rar& \so(2k{+}1)\oplus \so(2n-2k) \,, \non \\
\spl(2n) &\rar& \spl(2k)\oplus \spl(2n-2k) \,, \\
\so(2n) &\rar& \so(2k)\oplus \so(2n-2k) \,. \non 
\eea
It then follows that the rigid semisimple surface operators in the $B_n$ , $C_n$ and $D_n$ theories correspond to pairs of partitions in the following way \cite{Gukov:2008}. In the $B_n$ case a  rigid semisimple surface operator is labelled by a pair of partitions $(\la';\la'')$ where $\la'$ is a rigid  $B_k$ partition and  $\la''$ is a  rigid $D_{n-k}$ partition. For the $C_n$ theories a  rigid semisimple surface operator is labelled by a pair of partitions $(\la';\la'')$ where $\la'$ is a  rigid $C_k$ partition and  $\la''$ is a  rigid $C_{n-k}$ partition (and $k\ge \lfloor\frac{n}{2}\rfloor$, where $\lfloor\cdot\rfloor$ denotes the integer part). Finally, for the $D_n$ theories a  rigid semisimple surface operator is labelled by a pair of partitions $(\la';\la'')$ where $\la'$ is a rigid $D_k$ partition and  $\la''$ is a rigid $D_{n-k}$ partition (and $k\ge \lfloor\frac{n}{2}\rfloor$). (In all the above theories, the rigid unipotent surface operators arise as a limiting case when $\la''=0$.)

The Weyl group of a simple Lie group (algebra) is a finite group of particular importance. For the Weyl group corresponding to a classical group, both its conjugacy classes\footnote{Recall that a conjugacy class, $[h]$, comprises all elements obtained from $h$ by conjugation by a group element i.e. all elements of the form $g h g^{-1}$. Any element of the group belongs to precisely one conjugacy class. It is a known fact that any finite group has a certain number of conjugacy classes and an equal number of unitary representations. } and unitary representations  are in one-to-one correspondence with certain partitions. For the $A_n$ case both the set of conjugacy classes and the unitary representations are in one-to-one correspondence with the set of partitions of $n{+}1$. For the $B_n$ and $C_n$ theories (whose Weyl groups are isomorphic) both conjugacy classes and irreducible unitary representations are in one-to one correspondence with ordered pairs of partitions $[\al;\bet]$ where $\al$ is a partition of $n_\al$ and $\bet$ is a partition of $n_{\beta}$, such that $n_\al+n_\bet= n$. For the $D_n$ case there is also a correspondence with pairs of partitions $[\al;\bet]$ where again $n_\al+n_\bet= n$. However, in this case there are some further refinements, but as these will not play a role in this paper we will not describe them here.  
Finally, we mention that even though the conjugacy classes and unitary representations are parameterised by the same set of elements there is no canonical isomorphism between the two sets (except for the $A_n$ case).

There exist relations (maps) between the unipotent conjugacy classes (nilpotent orbits) of a simple group and its Weyl group. 
The Kazhdan-Lusztig map is a  (in general non-bijective) map from the unipotent conjugacy classes to the set of conjugacy classes of the Weyl group. 
The Springer correspondence is a (injective) map from the unipotent conjugacy classes  to the set of unitary representations of the Weyl group.
For the classical groups these maps can be described explicitly in terms of partitions. 
The simplest case is $A_n$ for which both the Kazhdan-Lusztig map and the Springer correspondence are given by the identity map. 

The  Kazhdan-Lusztig map can be extended to the case of rigid semisimple conjugacy classes using a result due to Spaltenstein~\cite{Spaltenstein:1992}. (As the  Kazhdan-Lusztig map for the  unipotent conjugacy classes is a special case of this construction we will not describe it separately.) Recall from the above discussion that the  rigid semisimple conjugacy classes are described by pairs of partitions $(\la';\la'')$ and that the conjugacy classes of the Weyl group are described by pairs of partitions $[\al;\beta]$. What is needed is therefore a map between these two classes of objects. Such a map can be explicitly constructed as follows. Start by adding the two partitions: $\la=\la'+\la''$. Then form the symplectic partition $\mu =Sp(\la)$ where the function $Sp$ is defined as follows. The parts of $\mu =Sp(\la)$ are given by
\be \label{Sp}
\mu_i = Sp(\la)_i = \left\{ \ba{ll} \la_i + p_{\la}(i) &\qquad \mbox{if $\la_i$ is odd and $\la_{i}\neq \la_{i-p_{\la}(i)}$ }\,, \\ \la_i &\qquad\mbox{otherwise}\,. \ea   \right.
\ee
where $p_\la(i) = (-1)^{\sum_{k=1}^{i}\la_k}$.
 The effect of this operation is to ensure that the odd parts of the resulting partition never occur an odd number of times, i.e.~the resulting partition is symplectic. As an example, if $\la = 7\,6^2\,5^3\,2^2\,1$ then $Sp(\la)=6^4\,5^2\,2^2$.

The next step is to define the function $\tau$ from the positive integers to $\pm 1$ in the following way. For the $B_n$ and $D_n$ cases $\tau(m)$ is $-1$ if $m$ is even and there exists at least one $\mu_i$ such that $\mu_i=m$ and either of the following three conditions is satisfied 
\be \label{tau}
\ba{ll}
(i) &\qquad \mu_i \neq \la_i  \,,  \\[3 pt]
(ii) &\qquad \sum_{k=1}^{i}\mu_k \neq   \sum_{k=1}^{i}\la_k  \,,\\[3 pt]
(iii)_{\SO} &\qquad \la'_i \; \mbox{  is odd}\,.
\ea
\ee
In all other instances $\tau$ is $1$. For $C_n$ the definition is the same except that condition $(iii)_{\SO}$ is replaced by 
\be \label{tau3sp}
\!\!\!\!\!\!\!\!\!\!\!\!\!\!\!\!\!\!\!\!\!\!\!\!\!\!\!\!\!\!
(iii)_{\Sp} \qquad \la'_i \; \mbox{  is even}\,.
\ee
Finally construct a pair of partitions $[\al;\beta]$ as follows. For each pair of parts of $\mu$ both equal to $a$ and such that $\tau(a)=1$ retain one part $a$. From the integers so obtained form the partition $\al$. For each part of $\mu$ of size $2b$ such that $\tau(2b)=-1$ retain $b$. From the integers so obtained form the partition $\beta$. The resulting pair of partitions $[\al;\beta]$ corresponds to  a conjugacy class of the Weyl group. See \cite{Spaltenstein:1992} for more details. As an example, $(\la';\la'')=(3\,2^2\,1^4;3\,2^2\,1^3)$ is mapped to $[\al;\bet]=[4\,;3\,1^3]$.

To describe the Springer correspondence for the classical groups it is convenient to use certain symbols introduced by Lusztig. This construction is described in chapter 10 of \cite{Collingwood:1993}. We briefly recall the main results here. 

In the $B_n$ case start by adding $l-k$ (where $l$ is the length of the partition) to the $k$th part of the partition. Then split the result into two sets: one containing the even parts and one containing the odd parts.   Arrange the odd parts in an increasing sequence and write them as $2f_i{+}1$ (starting with $f_1$). Similarly, write the even parts as $2g_i$ and  arrange them in an increasing sequence (starting with $g_1$). Next form $\al_i = f_i-i+1$ and $\bet_i = g_i-i+1$. Note that the number of $\al_i$'s is always one more than the number of $\bet_i$'s.  We then write the {\it symbol} as
\be \label{symbol}
\left(\ba{@{}c@{}c@{}c@{}c@{}c@{}c@{}c@{}} \al_1&&\al_2&&\al_3&&\cdots \\ & \bet_1 && \bet_2 && \cdots & \ea \right).
\ee
An example illustrates the method. The $B_{10}$ partition $\la=3^3\,2^4\,1^4$ has the symbol
\be \label{exs}
\left(\ba{@{}c@{}c@{}c@{}c@{}c@{}c@{}c@{}c@{}c@{}c@{}c@{}} 0&&0&&1&&1&&1&&1 \\ & 1 && 1 && 1 &&1&&2& \ea \right).
\ee
Viewing the two rows of the symbol as two partitions gives the Springer correspondence, since the resulting pair of partitions corresponds to a unitary representation of the Weyl group.

For the $C_n$ theory the symbol is formed in an analogous way. If the length of the partition is even, first append an extra $0$ as the last part of the partition; if the length is odd leave the partition unchanged. Then construct $f_i$ and $g_i$ as in the $B_n$ case and form $\al_i = g_i-i+1$ and $\bet_i = f_i-i+1$. The number of $\al_i$ is again one more than the number of $\bet_i$ and the symbol is written as in (\ref{symbol}). As an example the $C_{10}$ partition $\la=3^2\,2^6\,1^2$ has the symbol (\ref{exs}). 

For the $D_n$ theory one forms $f_i$ and $g_i$ exactly as in the $B_n$ case. The difference as compared to the $B_n$ case is that now the number of $f_i$ and $g_i$ are  equal. This means that there are two ways to write the symbol. For reasons that will become clear we use the definition $\al_i = g_i-i+1$ and $\bet_i = f_i-i+1$, i.e.~the opposite rule compared to the $B_n$ case. Conventionally one writes the symbol with two rows of equal length. However, since we are only interested in rigid partitions which always have at least one part equal to 1 and hence $\bet_1=0$ we will omit this entry  (and relabel $\bet_2\rar\bet_1$ etc.) when writing the symbol to ensure that the number of $\al_i$ is one more than the number of $\bet_i$ just as in the $B_n$ and $C_n$ cases.  As an example the rigid $D_{10}$ partition $\la=4^2\,3\,2^2\,1^5$ then has the symbol
\be 
\left(\ba{@{}c@{}c@{}c@{}c@{}c@{}c@{}c@{}c@{}c@{}} 1&&1&&2&&2&&2 \\ & 0 && 0 && 0 &&2& \ea \right).
\ee

As mentioned above the map provided by the Springer correspondence is only injective. There exists a way to extend it to a bijection. We will not describe this  extended Springer correspondence here as the relevance (if any) to surface operators is not clear.

The symbols as defined above provide an alternative characterisation of special partitions/surface operators. In the $B_n$ and $C_n$ theories a symbol is special if $\al_1\leq \bet_1\leq \al_2+1\leq \bet_2+1 \leq \cdots$. (The rigidity restriction can also be translated into the language of symbols.) In the $D_n$ theory a rigid symbol (defined as above) is special if  $\al_1\leq \bet_1+1\leq \al_2+1\leq \bet_2+2 \leq \cdots$.

A generalisation of the Springer correspondence to rigid semisimple conjugacy classes will be discussed in the following subsection.

\subsection{Invariants of surface operators: dimension, fingerprints and symbols} \label{inv} 
To investigate how the S-duality map acts on rigid surface operators it is very helpful to find invariants of the surface operators, i.e.~expressions which do not change under the S-duality map. In \cite{Gukov:2008} it was pointed out that the most basic invariant of a (rigid) surface operator is the dimension, $d$, of the associated orbit. This quantity is calculated as follows \cite{Gukov:2008,Collingwood:1993}:
\bea
B_n: & d = 2n^2 + n -\half \sum_{k} (s_k')^2 -  \half \sum_{k} (s_k'')^2 
+ \half \sum_{k\;\mathrm{odd}} r_k'+ \half \sum_{k\;\mathrm{odd}} r_k'' \,,\non \\
C_n: & d = 2n^2 + n -\half \sum_{k} (s_k')^2 -  \half \sum_{k} (s_k'')^2 
- \half \sum_{k\;\mathrm{odd}} r_k'- \half \sum_{k\;\mathrm{odd}} r_k''\,, \\
D_n: & d =2n^2 - n -\half \sum_{k} (s_k')^2 -  \half \sum_{k} (s_k'')^2 \non
+ \half \sum_{k\;\mathrm{odd}} r_k'+ \half \sum_{k\;\mathrm{odd}} r_k''\,.
\eea
Here $s'_k$ denotes the number of parts of $\la'$'s that are larger than or equal to $k$ and $r_k'$ denotes the number of parts of $\la'$ that are equal to $k$. The definitions of $s_k''$ and $r_k''$ are the same with respect to $\la''$.

In \cite{Gukov:2008} another more refined invariant was also constructed. This invariant arose by considering the singular behaviour of the fields near $r=0$. It was shown that the mathematical description of this invariant is precisely the Weyl group conjugacy class associated with the surface operator via the Kazhdan-Lusztig map. This means that the pair of partitions $[\al;\bet]$ constructed from $(\la';\la'')$ as in the previous subsection should not change under S-duality. In \cite{Gukov:2008} the Weyl group conjugacy class  arising from the Kazhdan-Lusztig map was referred to as the {\it fingerprint} of the surface operator; we will use this terminology throughout.

We now propose another invariant of rigid surface operators. This invariant is similar to the fingerprints but is based on the Springer correspondence rather than on the Kazhdan-Lusztig map.

The proposed invariant involves an extension of the Springer correspondence to rigid semisimple conjugacy classes and is constructed as follows (a similar construction appears in \cite{Hezard:2004}). Calculate the symbols for both $\la'$ and $\la''$ using the prescriptions given in the previous subsection and then add the two results `from the right', i.e.~write the symbols right adjusted and simply add the entries that are `in the same place'.  An example illustrates the addition rule:
\be \label{symboladd}
\left(\ba{@{}c@{}c@{}c@{}c@{}c@{}c@{}c@{}c@{}c@{}c@{}c@{}c@{}c@{}} 0&&0&&0&&0&&0&&1&&1 \\ & 1 && 1 && 1 &&1&&1&&2 & \ea \right) +
 \left(\ba{@{}c@{}c@{}c@{}c@{}c@{}c@{}c@{}c@{}c@{}c@{}c@{}} 0&&0&&0&&1&&1&&1 \\ & 1 && 1 &&1&&1&&1 & \ea \right)=
\left(\ba{@{}c@{}c@{}c@{}c@{}c@{}c@{}c@{}c@{}c@{}c@{}c@{}c@{}c@{}} 0&&0&&0&&0&&1&&2&&2 \\ & 1 && 2 && 2 &&2&&2&&3 & \ea \right).
\ee
We refer to the resulting expression as the {\it symbol} of the surface operator. 

It turns out that the symbol of a rigid surface operator contains the same amount of information as the fingerprint in the sense that if two rigid surface operators have the same symbols they also have the same fingerprints and vice versa. (We have not rigorously proven this statement but we have checked it in many cases.) The fact that the symbol is not an essentially new invariant is perhaps a bit disappointing but there are certain advantages of the symbols compared to the fingerprints since they are easier to calculate and their properties were quite useful in finding the S-duality maps we propose in later sections.
In particular, if one want to find all possible duals of a certain (rigid) surface operator one simply looks at all possible ways of splitting the corresponding symbol into two (rigid) symbols in the dual theory. There is always only a finite number of possibilities.

\subsection{Invariants of surface operators: centre and topology}
\label{ctinv}
In \cite{Gukov:2008} further discrete invariants were also constructed. We briefly recall the definitions here.  
Given a surface operator corresponding to some $V$ one can form $\ze \,V$ where $\ze$ is a non-trivial element of the centre of the gauge group. If these two expressions correspond  to two different surface operators then in the terminology of \cite{Gukov:2008} one says that the surface operator can detect the centre. However, if one can find a group element $g$ such that $g V g^{-1} = \ze\, V$ then $V$ and $\ze\,V$ belong to the same conjugacy class and do not correspond to different surface operators. 

Unipotent (rigid) surface operators can always detect the centre \cite{Gukov:2008}. For rigid semisimple surface operators the situation is more involved. In the $B_n$ case we should consider the gauge group $\Spin(2n{+}1)$ with centre $\ZZ_2$ generated by $-1$. Since both the rigid partitions $\la'$ and $\la''$ have at least one part equal to 1 (which corresponds to the trivial one-dimensional $\su(2)$ representation) then in the projection to the $\SO(2n{+}1)$ theory $V$ takes the form 
\be
\left(\ba{cc|cc} 1 & \cdots & & \\ \vdots &\ddots && \\ \hline && -1 &\cdots \\ && \vdots & \ddots \ea \right).
\ee
Now this matrix lifts to $V =  \ga_2 f(\ga_3,\ldots,\ga_{2n})$ in the $\Spin(2n{+}1)$ theory, where $\ga_i$ are the usual gamma matrices: $\{\ga_i,\ga_j\}=2\de_{ij}$.  This follows from the lifting
\be
\mathrm{O}(2) \ni \left(\ba{cr} 1&0\\0&-1\ea\right) \rar \ga_2 \in \Pin(2) \,.
\ee
If we then do a gauge rotation with $g=\ga_1\ga_2$ we find $g V g^{-1} = - V$. This means that rigid  semisimple surface operators  can never detect the centre  in the $B_n$ theory. 

In the $C_n$ theory (i.e.~$\Sp(2n)$ with centre $\ZZ_2$ generated by $-1$), we note that if $\la'$ and $\la''$ have an odd-dimensional part (or a pair of even-dimensional parts) in common, then these correspond to a block
\be \label{t+}
\left( \ba{cc} t_+ & 0 \\ 0& -t_+ \ea \right) \equiv t_+\otimes \si_z\,,
\ee
where $t_+$ belongs to a single odd-dimensional $\su(2)$ representation (or a sum of even-dimensional $\su(2)$ representations). In this case the symplectic unit acts inside $t_+$ and does not affect the $2\times 2$ block structure. If we then do a gauge rotation which in the relevant sector looks like $g = \id \otimes \si_x$ we find that the above block (\ref{t+})  gets multiplied by $-1$. Repeating this argument we see that if $\la'=\la''$ we can find a group element such that $g V g^{-1} = -V$, which means that such surface operators can not detect the centre. However, it appears that there are additional semisimple surface operators which can not detect the centre\footnote{\label{puzzle}If true, this fact will lead to some puzzles in later sections; we therefore suspect that there is a fault in the reasoning.}. For the above argument to go through it  looks to be sufficient that the number of times a  given odd-dimensional representation (or pair of even-dimensional representations) appear in the two semisimple factors are equal~$\!\!\!\mod 2$ (subject also to the condition that the representation(s) can not appear in only one of the two semisimple factors). Surface operators in the $C_n$ theory  which fulfill this requirement seem not to be able to detect the centre. For instance, if an odd-dimensional irreducible representation appears three times in the first factor and once in the second we get a diagonal matrix similar to  (\ref{t+}) but with $t_+$ appearing three times and $-t_+$ once along the diagonal. If we then  perform the above gauge rotation in each of the the three $2\times2$ subblocks containing $-t_+$ and one of the three $t_+$ we find that the diagonal $4\times4$ matrix gets multiplied by an overall $-1$. 

In addition to the above construction based on the centre, a related  quantity was also introduced in \cite{Gukov:2008}. This `topology' quantity involves the homology groups $\pi_1(H)$ and $\pi_1(G)$ (where $G$ is the gauge group and $H$ is the subgroup of $G$ left unbroken by $V$) rather than the centre. We will not describe the construction here (see \cite{Gukov:2008} for details); instead we only give the criterion for when a surface operator can `detect topology'. In the $B_n$ and $C_n$ theories the `detects/does not detect topology' property is a $\ZZ_2$ quantum number just like the  `detects/does not detect the centre' property is.

In the $B_n$ theory a surface operator can detect topology provided the corresponding partitions $\la'$ and $\la''$ are not both rather odd \cite{Gukov:2008}. In the $C_n$ theory a surface operator can detect topology provided the corresponding partitions $\la'$ and $\la''$ are both special \cite{Gukov:2008}. (Note the relation with the table in Corollary 6.1.6 in \cite{Collingwood:1993}.)

In \cite{Gukov:2008} it was argued that the two discrete quantum numbers discussed above should be interchanged under S-duality, so that if a surface operator can detect topology then its dual should detect the centre and vice versa.

\setcounter{equation}{0}
\section{Rigid surface operators in the $B_n$/$C_n$ theories}\label{BC}

In this section we discuss the  $B_n$ and $C_n$ theories and try to obtain information about the S-duality map between the rigid surface operators in these two theories. 

\subsection{Generating functions}
\label{genf}
Generating functions proved to be very useful in the works~\cite{Henningson:2007a,Wyllard:2007}. The generating functions for the total number of rigid surface operators clearly contains less information than an  explicit S-duality map acting on the rigid surface operators, but they could still prove to be important as a testing ground in the search for the exact map. We therefore start with a discussion of the generating functions. In the formul\ae{} below we use the notation
\be
(a,q)_k := \prod_{n=0}^{k-1} (1-aq^n)\,.
\ee
The total number of rigid unipotent operators in the $\SO(n)$ theory is given by the coefficient in front of $q^{n}$ in (the extra 1 is added for later convenience)
\bea \label{f}
&&   \!\!\!  \!\!\!\!\!\!\!\! 1+ \sum_{k=1}^{\infty} \Bigg[\sum_{\stackrel{i_1=1}{\scriptscriptstyle i_1\neq 2} }^{\infty} q^{i_1} 
\sum_{i_2=1 }^{\infty} q^{2 2i_2} \cdots \sum_{i_{2k-2}=1}^{\infty} q^{2 (2k-2)i_{2k-2}} \Big(\sum_{\stackrel{i_{2k-1}=1}{\scriptscriptstyle i_{2k-1}\neq 2}}^{\infty} q^{(2k-1)i_{2k-1}}  + \sum_{i_{2k}}^{\infty} q^{2 (2k) i_{2k}} \Big) \Bigg] \non \\
&=& 1+\sum_{k=1}^{\infty} \frac{q^{3k^2 - 2k}(-q^3;q^6)_k}{(q^2;q^2)_{2k}} \equiv f(q)\,.
\eea
Similarly, the total number of rigid unipotent operators in the $\Sp(2n)$ theory is given by the coefficient in front of $q^{2n}$ in (again we added an extra 1)
\bea \label{g}
&& \!\!\!\! \!\!\!\!\!\!1+ \sum_{k=1}^{\infty} \Bigg[\sum_{i_1=1 }^{\infty} q^{2 i_1} 
\sum_{\stackrel{i_2=1}{\scriptscriptstyle i_2\neq 2} }^{\infty} q^{2 i_2} \cdots 
\sum_{\stackrel{i_{2k-2}=1}{\scriptscriptstyle i_{2k-2}\neq 2}}^{\infty} q^{(2k-2)i_{2k-2}} 
\Big(\sum_{i_{2k-1}=1}^{\infty} q^{2(2k-1)i_{2k-1}}  
+ \sum_{\stackrel{i_{2k}=1}{\scriptscriptstyle i_{2k}\neq 2}}^{\infty} q^{2k i_{2k} } \Big) \Bigg] \non \\
&=& 1+ \sum_{k=1}^{\infty} \frac{q^{3k^2 - k}(1-q^{4k}+q^{6k})(-q^6;q^6)_k}{(1-q^{2k}+q^{4k})(q^2;q^2)_{2k}} \equiv g(q) \,.
\eea
Using the result (\ref{f}), the generating function for the total number of rigid surface operators (both unipotent and semisimple) in the $B_n$ theories becomes
\be \label{Brig}
 [f(q)^2-f(-q)^2 ]/4 \,.
\ee
Similarly in the $C_n$ case we find using (\ref{g}) the following generating function for the total number of rigid surface operators (here we multiplied the result by an extra factor of $q$ to facilitate the comparison with the $B_n$ result)
\be \label{Crig}
q\,[g(q)^2 + g(q^2)]/2 \,.
\ee
By expanding the above two expressions (\ref{Brig}) and (\ref{Crig}) one finds that the difference is
\be
q^9 + 2 \,q^{11} + 4\, q^{13} + 5\, q^{15} + 9\, q^{17} + 12 \,q^{19} + 17\, q^{21} + 23\, q^{23}+ \ldots
\ee
and hence there is a discrepancy between the number of rigid surface operators in the $B_n$ and $C_n$ theories. This discrepancy was first observed in the $B_4$/$C_4$ theories in \cite{Gukov:2008}. From the above expressions one gets some further insight into the discrepancy. It appears that (for $n\geq 4$) the number of rigid surface operators is larger in the $B_n$ theory as compared to the $C_n$ theory and that the excess grows with the rank, $n$. However,  the excess number of states divided by  the total number  appears to approach zero as $n\rar\infty$. This leads to the hope that only a minor modification is needed to make the numbers match. This dovetails nicely with the fact that most rigid surface operators do seem to have candidate duals. The discrepancy is clearly a major problem but we will ignore it for now and try to identify certain subsets of rigid surface operators and make proposals for how the S-duality map should acts on these.  We will return to the discrepancy issue in section \ref{discrep}.

\subsection{S-duality map between rigid special unipotent surface operators \cite{Gukov:2008} }

In \cite{Gukov:2008} it was proposed that the special rigid unipotent surface operators in the $B_n$ and $C_n$ theories are related by S-duality. As discussed above, special rigid unipotent surface operators in the $B_n$ theories are characterised by Young tableaux where all the rows have an odd number of boxes and the number of rows is also odd. (The tableaux of course also satisfy the conditions required for them to be rigid.)
Special rigid unipotent surface operators in the $C_n$ theories are described by Young tableaux where all the rows have an even number of boxes (plus the rigidity conditions).

The proposed S-duality map (which we will call $X_S$) from the special rigid unipotent surface operators in the $B_n$ theory to those in the $C_n$ theory acts in the following way \cite{Gukov:2008}
\bea \label{XS}
X_S:&& m^{2n_m+1}\, (m-1)^{2n_{m-1}}\, (m-2)^{2n_{m-2}  } \cdots 2^{n_2} \, 1^{2n_1}  \non \\   & \mapsto&
m^{2n_m}\, (m-1)^{2n_{m-1} +2 }\, (m-2)^{2n_{m-2} - 2 } \cdots 2^{n_2+2} \, 1^{2n_1-2}\,.
\eea
Here $m$ has to be odd in order for the first object to be a $B_n$ partition. Furthermore, it is clear that the map is a bijection so that $X_S^{-1}$ is well defined. 

The map (\ref{XS}) preserves the rigidity conditions since $n_{2j+1}\neq 1$ on the $B_n$ side implies  $n_{2j}\neq 1$ on the $C_n$ side.
Note that the map  (\ref{XS}) is essentially the `$p_C$ collapse' described in chapter 6.3  in \cite{Collingwood:1993} or more precisely the map $Sp$ described above and in \cite{Spaltenstein:1992}.  The inverse operation, $X_S^{-1}$, is essentially the `$p^B$ expansion' also described in chapter 6.3  in \cite{Collingwood:1993}.

The matching of the generating functions for the special unipotent surface operators in the $B_n$ and $C_n$ theories is the equality:
\be
\sum_{k=1}^{\infty} \frac{q^{6k^2 - 8k + 3}}{(q^2;q^2)_{2k-1}} = q + \sum_{k=1}^{\infty} \frac{q^{6k^2 - 4k + 1}(1-q^{4k}+q^{8k})}{(q^2;q^2)_{2k}} \,.
\ee

 In \cite{Gukov:2008} it was checked that the fingerprints and discrete invariants are preserved by the map.  On both sides the fingerprints become
\be
[\cdots 5^{n_5-1} \, 3^{n_3-1} \,  1^{n_1-1} \,  ; \cdots 2^{2n_4+2} \,  1^{2n_2+2}]\,.
\ee
On the $B_n$ side rigid special unipotent surface operators can detect the centre and the topology. The same is true on the $C_n$ side. 

Above we proposed an alternative invariant based on symbols. This invariant can be calculated on both sides and gives:
\be \label{spesym}
 \Bigg(\!\!\!\ba{c}  0\cdots 0\;\; \overbrace{1\cdots 1}^{n_2} \;\; 1 \dots 1 \;\; \cdots  \\ 
\;\;\;\; \underbrace{1 \cdots 1}_{n_1} \;\; 1\cdots 1 \;\;  \underbrace{2 \cdots 2}_{n_3}\;\;\cdots  \ea \!\!\!\Bigg). 
\ee
Note that the jumps in the entries occur on different rows each time. This alternating behaviour is characteristic of unipotent special surface operators.

It is not entirely obvious that the S-duality map (\ref{XS}) is uniquely fixed by the requirement that it preserves the invariants. Nevertheless, it is a simple rule and we will assume that it is the correct map.

\subsection{S-duality map for rigid rather odd unipotent surface operators }
\label{rodd}
Above we saw that the special unipotent operators are related by S-duality. In this subsection we will discuss another subclass of operators in the $B_n$ theories and identify their duals. This subclass consists of all $B_n$ operators for which one can detect the centre but not the topology. From the discussion in section \ref{ctinv} we find that surface operators with these properties are rigid rather odd unipotent surface operators.
Such surface operators correspond to partitions of the form $ \cdots 5 \,4^{2n_4}\, 3 \,2^{2n_2}\, 1 $ (note that the number of odd integers has to be odd for the surface operator to belong to $B_n$). 

We propose the duality map
\bea \label{oddmap}
 &&(\cdots  9\, 8^{2n_8}\, 7 \, 6^{2 n_6}  5\, 4^{2n_4}\, 3 \,2^{2n_2} \,1 \,; \emptyset ) \non \\ &\mapsto& (\cdots 4^{2n_8+2} 3^{2n_6} 2^{2n_4+2} 1^{2n_2} \, ; \cdots 4^{2n_8+2} 3^{2n_6} 2^{2n_4+2} 1^{2n_2} )\,.
\eea
We first note that the proposed duals are rigid (including the constraint that even parts can not appear with multiplicity 2.  Furthermore, the duals are special semisimple surface operators constructed out of two equal partitions. Since the surface operators are special they can detect the topology and since $\la'=\la''$ they can not detect the centre as required (cf.~section \ref{ctinv}).
Next one can easily calculate the fingerprints on both sides to obtain
\be
[\cdots6^{n_6}\,2^{n_2} \,; \cdots 4^{2n_8+2} 2^{2n_4+2} ]\,.
\ee
The matching of symbols can also be checked:
\bea
&&\Bigg(\!\!\!\ba{c} 0 \;\; 0\cdots 0 \;\; \overbrace{2\cdots 2}^{n_4+1} \;\; 2 \dots 2 \;\; \cdots  \\ 
\, \underbrace{2 \cdots 2}_{n_2} \;\; 2\cdots 2 \;\;  \underbrace{4 \cdots 4}_{n_6}\;\;\cdots  \ea \!\!\!\Bigg) =  \\
&&\Bigg(\!\!\!\ba{c} 0 \;\; 0\cdots 0 \;\; \overbrace{1\cdots 1}^{n_4+1} \;\; 1 \dots 1 \;\; \cdots  \\ 
\, \underbrace{1 \cdots 1}_{n_2} \;\; 1\cdots 1 \;\;  \underbrace{2 \cdots 2}_{n_6}\;\;\cdots  \ea \!\!\!\Bigg)
+ \Bigg(\!\!\!\ba{c} 0 \;\; 0\cdots 0 \;\; \overbrace{1\cdots 1}^{n_4+1} \;\; 1 \dots 1 \;\; \cdots  \\ 
\, \underbrace{1 \cdots 1}_{n_2} \;\; 1\cdots 1 \;\;  \underbrace{2 \cdots 2}_{n_6}\;\;\cdots  \ea \!\!\!\Bigg). \non
\eea
Thus the proposed dual pair passes all consistency checks that we know of.

The check of the matching of symbols is particularly revealing. This is because the symbol on the $B_n$ side only involves even numbers, and jumps alternate between the two rows.  There is only {\it one} way to write it as a sum of two rigid special $C_n$ symbols (recall from the above discussion, cf.~(\ref{spesym}), that rigid special $C_n$ symbols also have jumps alternating between the two rows but the jumps only involve a difference of $+1$ each time).  The fact that the surface operators need to detect topology on the $C_n$ side (since the centre can be detected on the $B_n$ side), requires the $C_n$ partitions to be special and we can be confident that we have found the right dual. 
 For this reason the class of rigid rather odd unipotent operators is in a sense even simpler that the class of special unipotent operators whose duals where identified in \cite{Gukov:2008} and described above. 

The matching of the generating functions of the two dual classes is the equality:
\be
 \sum_{k=1}^{\infty} \frac{q^{3k^2 - 2k }}{(q^4;q^4)_{k}} = q + \sum_{k=1}^{\infty} \frac{q^{12k^2 - 8k + 1}(1-q^{8k}+q^{16k})}{(q^4;q^4)_{2k}} \,.
\ee

Let us now describe how the map (\ref{oddmap}) acts on the partitions in a way which will facilitate the generalisation to all rigid unipotent  $B_n$ surface operators (not necessarily special or rather odd). For simplicity we focus on the case $5\,4^2\,3\,2^4\, 1$. The Young tableau is 
\be
\tableau{1 3 4 8 9} 
\ee
Let us split this tableau into one tableau constructed from the rows with an odd number of boxes and one tableau constructed from the rows with an even number of boxes; we get
\be
\tableau{1 3 9}\quad  ;\quad \tableau{4 8} 
\ee
Next we apply the map (\ref{XS}) to the tableau constructed from the odd rows to obtain 
\be
\tableau{4 8}\quad  ;\quad \tableau{4 8} 
\ee
Thus we arrive at  the rigid surface operator $(2^4\,1^4;2^4\,1^4)$ which agrees with the dual proposed above. The general prescription is clear: split the Young tableau into one tableau constructed from the rows with an odd number of boxes and one constructed from the rows with an even number of boxes, and then apply the map (\ref{XS}) to the tableau with odd rows. The two tableaux so constructed correspond the two (equal) rigid special symplectic partitions given in (\ref{oddmap}).

\subsection{A proposal for the S-duality map for rigid unipotent operators }
\label{unip}
From the discussion in the previous two subsections, a natural generalisation of the proposed S-duality map to all unipotent surface operators in the $B_n$ theories now presents itself: we simply apply the same manipulation rule at the level of Young tableaux that we did at the end of the previous subsection. 
First note that this algorithm always gives a rigid special semisimple surface operator in the $C_n$  theory: The first tableaux (the one with only odd rows) is always a special rigid partition in some $B_k$ theory and the map (\ref{XS}) turns this into a special partition in the $C_k$ theory. The second partition (the one with only even rows) already corresponds to a special rigid partition in the $C_{n-k}$ theory and is left untouched. The fact that we obtain a special $C_n$ surface operator shows that it can detect topology on the $C_n$ side which is consistent with the fact that the unipotent $B_n$ surface operators can detect the centre.  Also note that if the $B_n$ tableaux is special (i.e.~has only odd rows) we recover the map proposed in \cite{Gukov:2008}. 

As an example of the procedure, consider the unipotent $B_{16}$ operator corresponding to the partition $\la = 5\, 4^2 \, 3^3\, 2^4\, 1^3$. Applying the proposed map we find
\be \label{rank16}
\tableau{1 3 6 10 13} \quad\mapsto \quad\tableau{4 12}\quad  ;\quad \tableau{6 10} 
\ee
i.e. $(2^4\,1^8\,,\,2^6\,1^4)$. This semisimple $C_{16}$ operator has identical fingerprint and symbol to the $B_{16}$ operator we started with.  

To check that the proposed map preserves the symbols is not too difficult. 
This is because the Young tableau operations we have performed have direct counterparts in the symbols. Splitting a rigid Young tableau along rows corresponds to splitting the symbol at the places where the values of the entries jump. In particular, splitting the Young tableau into sets of even and odd rows 
corresponds to splitting the symbol into two special symbols. To see this first note that there can be at most two consecutive jumps within either of the rows of the symbol before the jump has to switch rows (this follows from the orthogonal/symplectic constraint). The splitting into even and odd rows is  at points where the second of two such consequtive jumps occurs.  
This is best illustrated by an example. In the above rank 16 tableau (\ref{rank16}) the splitting of symbols is as in (\ref{symboladd}).

It is also possible to show that the proposed map preserves the fingerprints. This is a little more involved. 
The first thing to note is that on the $B_n$ side $\la=\la_{\mathrm{even}} + \la_{\mathrm{odd}}$ and $\mu = Sp(\la) = Sp(\la_{\mathrm{odd}})+\la_{\mathrm{even}}$. This result follows from the definition (\ref{Sp}). Note that the longest row in a rigid $B_n$ partition always contains an odd number of boxes. The following two rows are either both of odd length or both of even length. This pairwise pattern then continues. If the tableau has an even number of rows the row of shortest length has to be even.   

On the $C_n$ side $\la' = X_S(\la_{\mathrm{odd}}) = Sp(\la_{\mathrm{odd}})$ and $\la''=\la_{\mathrm{even}}$ which implies that $\mu=\la$ since both $\la'$ and $\la''$ are special $C_n$ partitions which means that so is $\la\equiv \la'+\la''$.
Since $\mu =\la$ it then follows from the definition of the map $\tau$ that $\tau$ is $-1$ only when $\mu_i$ is even and $\la'_i$ is even, i.e.~when both $\la'_i$ and $\la''_i$ are even. 

We need to show that $\tau$ is also $-1$ for the same $\mu_i$ on the $B_n$ side. When $\mu_i$ is even, either both of the corresponding parts of $Sp(\la_{\mathrm{odd}})$ and $\la_{\mathrm{even}}$ are odd or they are even. If both are odd we see 
 from (\ref{XS}) that the first two conditions in (\ref{tau}) are fulfilled (the third condition is moot when $\la''=\emp$). Hence $\tau=+1$ for such $\mu_i$.
If both are even it follows from (\ref{XS}) that for the even parts of $Sp(\la_{\mathrm{odd}})$ at least one of the corresponding parts of $\la_{\mathrm{odd}}$ is different. This implies that we have $\tau=-1$. This is the same result as on the $C_n$ side, hence the fingerprints are the same.   

As already mentioned the fact that the $B_n$ unipotent surface operators detect the centre is consistent with the fact that the proposed duals detect topology. The  $B_n$ unipotent surface operators that detect topology (i.e.~the ones that are not rather odd) should have duals which detect the centre. Here we encounter a puzzle: from the discussion in section \ref{ctinv} it seems that some special rigid semisimple $C_n$ operators with $\la'\neq \la''$ do not detect the centre. 
If so, this would be problematic for our proposed map. This leads us to suspect, as was already mentioned in footnote \ref{puzzle}, that the arguments in section \ref{ctinv} are not completely correct. On the other hand, if the arguments are correct then we have a more severe problem since there are in many cases no other possible duals apart from the ones arising via our proposed map (for instance, this is the case for the surface operators with orbit-dimension 20 in the rank 6 example listed in the appendix). 
Another puzzling aspect of a similar nature is the following. In our proposal, the unipotent rigid $B_n$ surface operators get mapped into special rigid semisimple $C_n$ surface operators. But, the number of special rigid semisimple surface operators in the $C_n$ theories is larger than the number of rigid unipotent surface operators in the $B_n$ theories. This is problematic since we argued in section \ref{ctinv} that the special $C_n$ surface operators detect topology whereas the only $B_n$ surface operators which detect the centre are the unipotent ones. On the other hand based only on their fingerprints/symbols the extra rigid special $C_n$ surface operators appear to have candidate rigid $B_n$ duals.

Turning to the unipotent operators on the $C_n$ side we can make a similar proposal for the dual of these operators. Starting with the Young tableau corresponding to a rigid unipotent $C_n$ operator we split it into even-row and odd-row tableaux as in the $B_n$ case. (Note that the number of rows in the  odd-row tableau is always even.) We then apply the map $X_S^{-1}$ to the even-row tableau to obtain a $B_k$ tableau (this guarantees that we reproduce the map in \cite{Gukov:2008} for the special $C_n$ operators). From the odd-row tableaux  we want to obtain a rigid $D_{n-k}$ partition (since the operation on the even-row tableau gave us a $B_k$ partition).   
To accomplish this goal, we apply the following map 
\bea \label{YS}
Y_S:&& m^{2n_m+1}\, (m-1)^{2n_{m-1}}\, (m-2)^{2n_{m-2}  } \cdots 2^{n_2} \, 1^{2n_1}  \non \\   & \mapsto&
m^{2n_m}\, (m-1)^{2n_{m-1} +2 }\, (m-2)^{2n_{m-2} - 2 } \cdots 2^{n_2-2} \, 1^{2n_1+2}\,.
\eea
Here $m$ has to be even in order for the first element to be a $C_k$ partition. This map is very similar to the map (\ref{XS}) and takes a special $C_k$ partition to a special $D_k$ partition (note that the map preserves the number of boxes).
The map (\ref{YS}) is simply the `$p_D$ collapse' mentioned in chapter 6.3 in \cite{Collingwood:1993}. The inverse map, $Y_S^{-1}$, is the `$p^C$ expansion' (cf. chapter 6.3 in \cite{Collingwood:1993}).

 The proposed map therefore takes us from a rigid unipotent $C_n$ surface operator to a rigid special semisimple surface operator in the $B_n$ theory. Since such a surface operator on the $B_n$ side is never rather odd we can detect topology on the $B_n$ side which matches the fact that we can detect the centre on the $C_n$ side. Furthermore, the map maps unipotent $C_n$ surface operators which can also detect topology (the special unipotent surface operators) into $B_n$ surface operators which can also detect the centre (special unipotent surface operators).  The proposed  S-duality map for unipotent $C_n$ surface operators therefore does not suffer from the problems mentioned above for the map of $B_n$ unipotent surface operators, however, in the present case the map is less unique since there is no clear reason why the semisimple $B_n$ duals should be special.

Again one can check that the symbols match for the proposed dual pairs. The method is completely analogous to the one used for the unipotent $B_n$ surface operators so we will not repeat the details. 

To verify that the fingerprints also match we first note that the longest two rows in a rigid $C_n$ partition both contain  either an odd number or an even number of boxes. This pairwise pattern then continues. If the tableau has an odd number of rows the row of shortest length has to contain an even number of boxes. Since the unipotent $C_n$ partition is symplectic we have $\mu =Sp(\la) =\la$. It follows from this result that $\tau$ is $-1$ for all even $\mu_i$. From the above properties of rigid $C_n$ partitions, it also follows that the corresponding $\la_{\mathrm{even},i}$ and $\la_{\mathrm{odd},i}$ both have to be even.
On the $B_n$ side we have $\la = X_S^{-1} \la_{\mathrm{even}} + Y_S \la_{\mathrm{odd}}$ and $\mu =  \la_{\mathrm{even}} + \la_{\mathrm{odd}}$ (which follows from the definitions of $X_S$ and $Y_S$). As above, when $\mu_i$ is even we have that the corresponding $\la_{\mathrm{even},i}$ and $\la_{\mathrm{odd},i}$ both have to be even. When $\la_{\mathrm{even},i}$ is even there exists an $i$ such that $\mu_i$ and $\la_i$ differ, which means that $\tau$ is $-1$ for such $i$. This agrees with the $C_n$ result and the fingerprints are therefore the same.

We close this section by pointing out that 
in \cite{Lusztig:1984}, chapter 13.3, Lusztig constructs a map from unipotent (not necessarily rigid) $B_n$ [$C_n$] conjugacy classes to special (not necessarily rigid) $C_n$ [$B_n$] semisimple conjugacy classes. The map is not described in a very explicit way. However, in a later work \cite{Hezard:2004} a much more explicit map is constructed. The maps constructed in section 4.2 of  \cite{Hezard:2004} are very similar to the maps we have proposed. But, somewhat surprisingly, they are not the same maps since the maps in  \cite{Hezard:2004}, as far as we can see, do not preserve the rigidity conditions.

\subsection{A proposal for the S-duality map for $(\rho\,;\rho)$  $C_n$ surface operators }

Semisimple surface operators in the $C_n$ theories for which $\la'$ and $\la''$ are equal can not detect the centre (see section \ref{ctinv} above). Note that $n$ has to be even in order for such surface operators/partitions to exist. We argued above that the $(\rho\,;\rho)$ $C_n$ surface operators which are also special are dual to the rigid rather odd unipotent surface operators in the $B_n$ theory. We will now make a proposal for the dual of a general rigid  $(\rho\,;\rho)$ $C_n$ surface operator.  
Since such surface operators can detect neither centre nor topology  one expects the dual to be given by rigid rather odd semisimple operators in the $B_n$ theory since such operators have the same properties. 
 
Start by splitting the two equal tableaux into even-row and odd-row tableaux as above. Next apply the map (\ref{YS}) to one of the odd-row tableaux and apply the inverse of (\ref{XS}) to the even-row tableau in the other semisimple factor. Then add the altered and unaltered even-row tableaux to form one of the two partitions in a semisimple $B_n$ operator. Finally, do the same to the odd-row tableaux. In other words, the resulting $B_n$ partition becomes $(\rho_{\mathrm{even}} + X_S^{-1}\rho_{\mathrm{even}}\,;\rho_{\mathrm{odd}} + Y_S\rho_{\mathrm{odd}})$. Note that the first partition is a $B_k$ partition and the second factor is a $D_{n-k}$ partition.  As an example consider the $C_{14}$ operator $(4\, 3^2\, 2\, 1^2\,;\,4\, 3^2\, 2\, 1^2)$. Applying the suggested map we find:
\bea 
( \tableau{1 3 4 6} \; ; \;\tableau{1 3 4 6} ) &\mapsto& ( \tableau{4 6} + \tableau{1 3 7}  \; ;  \; \tableau{1 3} + \tableau{4}) \non \\
&=& ( \tableau{1 3 4 6 7} ; \tableau{1 3 4} )
\eea
i.e.~the  semisimple $B_{14}$ operator $(5\, 4^2\, 3\, 2^2\, 1\,;\,3\, 2^2\, 1)$ which is rather odd as expected.  Note that if the even-row tableaux  $\rho_{\mathrm{even}}$ is empty the inverse map (\ref{XS})  applied to it gives the partition 1.

To check that the symbols match one can use the same methods as in previous cases. The $C_n$ symbol corresponding to $\la=\rho+\rho$ has entries with only even numbers. This can be split into two symbols corresponding to rather odd symbols (for which the jumps are with steps of $+2$ and alternating between the rows, cf.~section \ref{rodd}) using the same methods as in section \ref{unip} except that now all entries are even. 

To verify that the fingerprints agree we start on the $C_n$ side where $\la = \rho+\rho$ and $\mu=\la$ since $\la$ is symplectic (it has only even parts). From this result it follows that $\tau$ is $-1$ whenever $\rho_i$ is even.

On the $B_n$ side we have  $\la = \rho + Y_S \rho_{\mathrm{odd}} + X_S^{-1} \rho_{\mathrm{even}}$ and $\mu = \rho+\rho$. When $\rho_i$ is even there exists $\mu_i$ which differ from the corresponding $\la_i$ and therefore $\tau$ is $-1$ for such $\mu_i$. When $\rho_i$ is odd one instead finds that $\tau$ is $1$. These results agree with the ones on the $C_n$ side and hence the fingerprints agree.

Thus the proposed dual pairs passes all consistency checks. However, we note that the number of rather odd semisimple $B_n$ surface operators is larger than the number of $(\rho\,;\rho)$ non-special surface operators on the $C_n$ side. 

\subsection{A proposal for the S-duality map for $(1;\de)$  $B_n$ surface operators }
Another class of surface operators for which a natural S-duality action exists are the rigid semisimple $B_n$ surface operators that are of the form  $(1;\de)$, i.e.~$\la'$ is a $B_0$ partition (1) and $\la''$ is a $D_n$ partition ($\de$). The proposed map is similar to the above examples: split the partition $\de$ into even and odd rows and leave the odd-row tableau unchanged and apply $Y_S^{-1}$ to the even-row tableau. Form a semisimple $C_n$ surface operator from the resulting two partitions. This operation gives a semisimple $C_n$ operator where both of the two partitions have only odd rows. 

Note that the above map is consistent with the proposed map for unipotent $C_n$ surface operators (when $\de$ is special and the dual unipotent $C_n$ operator has only odd rows) as well as with the map for $(\rho\,;\rho)$ $C_n$ surface operators  (when $\de$ is rather odd the dual has a $\rho$ with only odd rows).

The methods used to check that the dual pairs have the same fingerprints are similar to the previous cases.  Note that the longest row in a rigid $D_n$ partition always contains an even number of boxes. The following two rows are either both of odd length or both of even length. This pairwise pattern then continues. If the tableau has an even number of rows the row of shortest length has to be even. On the $B_n$ side $\la = 1 + \de_{\mathrm{odd}} + \de_{\mathrm{even}}$  and $\mu =\de_{\mathrm{odd}} +  Y_S^{-1} \de_{\mathrm{even}}$.  On the $C_n$ side $\la \equiv \la'+\la''= \de_{\mathrm{odd}} +  Y_S^{-1} \de_{\mathrm{even}}$ and $\mu = \la$.  This implies that whenever  $\de_{{\mathrm{odd}},i}$ is even $\tau$ is $-1$. This can be seen to be in agreement with the $B_n$ result (using the properties of the $Y_S$ map).

Excluding the case when $\de$ is rather odd, the fact that the surface operators on the $B_n$ side can detect topology means that  on the $C_n$ side the dual surface operators should detect the centre. Although this is generically the case, it seems that if the analysis in section \ref{ctinv} is correct some of the possible duals might not detect the centre. But as already mentioned in section \ref{unip} and footnote \ref{puzzle} we suspect that there are probably some misconceptions in that analysis. 

\subsection{General semisimple operators: search for an S-duality map}

Above we have made some proposals for how the S-duality map should act on certain subclasses of rigid surface operators. Our proposals include all unipotent rigid surface operators as well as certain subclasses of rigid semisimple operators. 
The goal is of course to extend the analysis to arbitrary rigid semisimple operators. 
 However, it seems that before such an extension can be found,  the reason for the mismatch of the total number of rigid surface operators in two theories must be resolved.  We therefore make some preliminary comments about the rigid surface operators responsible for the mismatch in the next subsection.

\subsection{Characterising the operators which seemingly have no dual} \label{discrep}

We saw in section \ref{genf} that there is an excess of rigid surface operators in the $B_n$ theories (when $n\geq 4$). 
One could speculate that it is only the excess surface operators which are problematic and which do not have duals, but this naive guess is not correct as we will see below. 
 
We will only attempt a preliminary analysis of which of the surface operators are problematic; our motivation is that a more thorough understanding of which surface operator do not have candidate duals might lead to progress. 
 
Our analysis will be based on the assumption that the symbols as defined in section \ref{inv} are invariants and we therefore start by recalling some pertinent facts. 
For rigid partitions of the form $\cdots 2^{2n_2}1$ in the $B_n$ theories the symbols take the form
\be
\left(\ba{@{}c@{}c@{}c@{}c@{}} 2&&\cdots \\ & 0 &&  \!\!\!\cdots  \ea \right),
\ee
whereas for rigid partitions of the form $\cdots 1^{n_1}$ with $n_1\ge 3$ the symbols take the form
\be
\left(\ba{@{}c@{}c@{}c@{}c@{}} 1&&\cdots \\ & 0 &&  \!\!\!\cdots  \ea \right).
\ee
Similarly in the $D_n$ theories one finds that the symbols take the forms
\be
\left(\ba{@{}c@{}c@{}c@{}c@{}} 0&&\cdots \\ & 2 &&  \!\!\!\cdots  \ea \right) \,,\qquad
\left(\ba{@{}c@{}c@{}c@{}c@{}} 0&&\cdots \\ & 1 &&  \!\!\!\cdots  \ea \right),
\ee
for rigid partitions of the form $\cdots 2^{2n_2}1$ and  $\cdots 1^{n_1}$ with $n_1\ge 3$, respectively.
In the $C_n$ theories the symbols take the form
\be \label{Cnsymbols}
\left(\ba{@{}c@{}c@{}c@{}c@{}} 1&&\cdots \\ & 0 &&  \!\!\!\cdots  \ea \right) \,,\qquad
\left(\ba{@{}c@{}c@{}c@{}c@{}} 0&&\cdots \\ & 1 &&  \!\!\!\cdots  \ea \right),
\ee
for rigid partitions of odd and even length, respectively.

Now consider semisimple surface operator in the  $B_n$ theory with symbols
\be \label{nodual}
\left(\ba{@{}c@{}c@{}c@{}c@{}} 2&&\cdots \\ & 2 &&  \!\!\!\cdots  \ea \right) \,,\qquad
\left(\ba{@{}c@{}c@{}c@{}c@{}} 2&&\cdots \\ & 1 &&  \!\!\!\cdots  \ea \right) \,,\qquad
\left(\ba{@{}c@{}c@{}c@{}c@{}} 1&&\cdots \\ & 2 &&  \!\!\!\cdots  \ea \right).
\ee
Surface operators with such symbols can {\it not} have (rigid) $C_n$ duals since in the $C_n$ theory such symbols can not be constructed from the sum of two symbols of the form (\ref{Cnsymbols}).
The above classes of $B_n$ operators (\ref{nodual}) correspond to pairs of partitions $(\la',\la'')$ where the length of $\la'$ is equal to the length of $\la''$ plus one, and one (or both) of $\la'$ and $\la''$ is of the form  $\cdots 2^{2n_2}1$.

There are further infinite classes of surface operators that can not have duals, e.g.~the $B_n$ ones that have symbols of the form 
\be 
\left(\ba{@{}c@{}c@{}c@{}c@{}c@{}} 1&&2&\; \cdots \\ & 1 && \cdots & \ea \right) \,,\qquad
\left(\ba{@{}c@{}c@{}c@{}c@{}c@{}} 0&&2&\; \cdots \\ & 1 && \cdots & \ea \right) \,,\qquad
\left(\ba{@{}c@{}c@{}c@{}c@{}c@{}} 0&&1&&\!\!\! \cdots \\ & 1 && 2 &\;\cdots \ea \right) .
\ee
We will not attempt to classify all symbols which can appear on the $B_n$ side but not on the $C_n$ side. Such a classification would anyway not be the end of the story since 
in addition to such symbols there are also symbols which can arise from two surface operators on the $B_n$ side but only from one on the $C_n$ side. This is a mismatch of a different type. Examples of such symbols include
\be
\left(\ba{@{}c@{}c@{}c@{}c@{}c@{}c@{}} 1&&1&&1\; \cdots \\ & 1 && 2 &\cdots \ea \right) \,,\qquad
\left(\ba{@{}c@{}c@{}c@{}c@{}c@{}c@{}c@{}c@{}} 1&&1&&1&&1\; \cdots \\ & 1 && 1&&2 &\cdots  \ea \right) .
\ee
The above examples have all been cases where there are too many $B_n$ surface operators of a certain type. Based on the generating functions a natural guess would have been that this would be the only type of problem. However, perhaps somewhat surprisingly, this is not true. Starting at rank 10 states appear in the $C_n$ theories which based on their symbols (and fingerprints) can not have duals in the $B_n$ theories. The first example in this series is
\be \label{c10}
( 2^4\,1^2\,;3^2\,2\,1^4)\,,
\ee
with symbol
\be
\left(\ba{@{}c@{}c@{}c@{}c@{}c@{}c@{}c@{}c@{}} 0&&1&&2 &&2 \\ & 1 && 1 && 3 & \ea \right).
\ee
Note that there appears to be a relation between (\ref{c10}) and  the excess problematic $B_4$ surface operator found in \cite{Gukov:2008}, namely $( 1^4\,;2^2\,1)$: the Young tableaux of this surface operator are obtained by removing the first rows in the two tableaux corresponding to the partitions in (\ref{c10}).

\setcounter{equation}{0}
\section{The $D_n$ theories} \label{D}

In this section we will very briefly discuss the extension of some of the techniques used in the $B_n$/$C_n$ theories to the $D_n$ (i.e.~$\SO(2n)$) theories. The discrete  invariants are potentially more restrictive since in this case the centre of $\Spin(2n)$ is of order 4, but we will not make use of them here. 

For unipotent $D_n$ operators we propose that the S-dual surface operator is obtained by splitting the corresponding tableau into even- and odd-row tableaux, applying the map $Y_S$ to the odd-row tableau (which corresponds to a $C_k$ partition) and leaving the even-row tableau unchanged. This operation results in a special semisimple rigid $D_n$ surface operator. One can check that the fingerprints and symbols are preserved by the map but we refrain from giving the details here. 

As another example consider semisimple rigid $D_n$ surface operators of the form $(\rho\,;\rho)$. We propose the following S-duality map. Split each $\rho$ into even- and odd-row tableaux and apply $Y_S$ to one of the odd-row tableau and $Y_S^{-1}$ to one of the even-row tableau. Then add the unchanged even-row tableau and the transformed even-row tableau and do the the same for the odd-row tableau. This procedure results in a rigid semisimple rather odd  $D_n$  surface operator.  Note that if $\rho$ is rather odd from the beginning then the proposed map leaves the surface operator unchanged.  Again one can check that the fingerprints and symbols are preserved by the proposed map. 

\setcounter{equation}{0} 
\section{Summary and open problems}
In this paper we have made some proposals for how the S-duality map should act on certain classes of rigid surface operators in the $B_n$, $C_n$ and $D_n$ theories. In particular, we have made proposals for all unipotent rigid surface operators as well as for some classes of rigid semisimple surface operators. Our proposed maps are speculative but their descriptions are quite simple and uniform. Attemts to continuing the analysis to more general classes of semisimple surface operators are hampered by the mismatch in the total number of rigid surface operators in the $B_n$ and $C_n$ theories. Since the $D_n$ theories are self-dual they might prove to be easier to study.  We took some tentative steps towards a classification of the $B_n$/$C_n$ rigid surface operators which can not have a dual, but the physical reason for the mismatch is still unknown.  Maybe the weakly rigid surface operators discussed in \cite{Gukov:2008} will play a role in the resolution. Clearly more work is required; hopefully  our constructions will be helpful in making further progress.

\section*{Acknowledgements}
The author wishes to thank M\aa ns Henningson for discussions.   
This work was supported by a grant from the Swedish Research Council.

\newpage

\appendix

\setcounter{equation}{0}
\section{Rigid surface operators in the $\SO(13)$ and $\Sp(12)$ theories}
Below we list (with no particular ordering) all rigid surface operators in the  $\Sp(12)$ and $\SO(13)$ theories. These tables illustrate the results in this paper. The first column lists the pair of partitions corresponding to the surface operator, the second column the dimension, the third the symbol, and the fourth the fingerprint.

\be \non {\small
\begin{array}{l@{\hspace{50pt}}l@{\hspace{50pt}}c@{\hspace{50pt}}l}
(1^{12}\,;\emp)
 & 
 0
 & 
\left( \begin{array}{@{}c@{}c@{}c@{}c@{}c@{}c@{}c@{}c@{}c@{}c@{}c@{}c@{}c@{}}
 0&&0&&0&&0&&0&&0&&0 \\
 &1&&1&&1&&1&&1&&1&
\end{array} \right)
 & 
[1^6\,;\emp]
 \\
 (2\,1^{10}\,;\emp)
 & 
 12
 & 
\left( \begin{array}{@{}c@{}c@{}c@{}c@{}c@{}c@{}c@{}c@{}c@{}c@{}c@{}}
 1&&1&&1&&1&&1&&1 \\
 &0&&0&&0&&0&&0&
\end{array} \right)
 & 
[1^5\,;1]
 \\
 (1^{10}\,;1^2)
 & 
 20
 &
\left( \begin{array}{@{}c@{}c@{}c@{}c@{}c@{}c@{}c@{}c@{}c@{}c@{}c@{}}
  0&&0&&0&&0&&0&&0 \\
 &1&&1&&1&&1&&2&
\end{array} \right)
 & 
[2\,1^4\,;\emp]
 \\
(2^3\,1^6\,;\emp) 
 & 
 30
 & 
\left( \begin{array}{@{}c@{}c@{}c@{}c@{}c@{}c@{}c@{}c@{}c@{}}
  1&&1&&1&&1&&1 \\
 &0&&0&&0&&1&
\end{array} \right)
 & 
[1^3\,;1^3]
 \\
 (2\,1^8\,;1^2)
 & 
 30
 & 
\left( \begin{array}{@{}c@{}c@{}c@{}c@{}c@{}c@{}c@{}c@{}c@{}}
  1&&1&&1&&1&&1 \\
 &0&&0&&0&&1&
\end{array} \right)
 & 
[1^3\,;1^3]
 \\
 (1^{8}\,;1^4)
 & 
 32
 & 
\left( \begin{array}{@{}c@{}c@{}c@{}c@{}c@{}c@{}c@{}c@{}c@{}}
  0&&0&&0&&0&&0 \\
 &1&&1&&2&&2&
\end{array} \right)
 & 
[2^2\,1^2\,;\emp]
\\
(2^4\,1^4\,;\emp)
 & 
 36
 & 
\left( \begin{array}{@{}c@{}c@{}c@{}c@{}c@{}c@{}c@{}c@{}c@{}}
  0&&0&&0&&1&&1 \\
 &1&&1&&1&&1&
\end{array} \right)
 & 
[1^2\,;1^4]
\\
 (1^8\,;2\,1^2)
 & 
 36
 & 
\left( \begin{array}{@{}c@{}c@{}c@{}c@{}c@{}c@{}c@{}c@{}c@{}}
  0&&0&&0&&1&&1 \\
 &1&&1&&1&&1&
\end{array} \right)
 & 
[1^2\,;1^4]
 \\
(1^6\,;1^6)
 & 
 36
 & 
\left( \begin{array}{@{}c@{}c@{}c@{}c@{}c@{}c@{}c@{}}
 0&&0&&0&&0 \\
 &2&&2&&2&
\end{array} \right)
 & 
[2^3\,;\emp]
 \\
(2^5\,1^2\,;\emp) 
 & 
 40
 & 
\left( \begin{array}{@{}c@{}c@{}c@{}c@{}c@{}c@{}c@{}}
 1&&1&&1&&1 \\
 &0&&1&&1&
\end{array} \right)
 & 
[1\,;1^5]
 \\
 (2\,1^6\,;1^4)
 & 
 40
 & 
\left( \begin{array}{@{}c@{}c@{}c@{}c@{}c@{}c@{}c@{}}
  1&&1&&1&&1 \\
 &0&&1&&1&
\end{array} \right)
 & 
[1\,;1^5]
 \\
 (1^6\,;2\,1^4)
 & 
 42
 & 
\left( \begin{array}{@{}c@{}c@{}c@{}c@{}c@{}c@{}c@{}}
  0&&1&&1&&1 \\
  &1&&1&&1&
\end{array} \right)
 & 
[\emp\,;1^6]
 \\
(3^2\,2\,1^4\,;\emp) 
 & 
 44
 & 
\left( \begin{array}{@{}c@{}c@{}c@{}c@{}c@{}c@{}c@{}}
  1&&1&&1&&1 \\
 &0&&0&&2&
\end{array} \right)
 &
[3\,1^2\,;1] 
 \\
 (2^3\,1^4\,;1^2)
 & 
 44
 & 
\left( \begin{array}{@{}c@{}c@{}c@{}c@{}c@{}c@{}c@{}}
  1&&1&&1&&1 \\
  &0&&0&&2&
\end{array} \right)
 & 
[3\,1^2\,;1]
 \\
 (2\,1^6\,;2\,1^2)
 & 
 44
 & 
\left( \begin{array}{@{}c@{}c@{}c@{}c@{}c@{}c@{}c@{}}
  1&&1&&2&&2 \\
 &0&&0&&0&
\end{array} \right)
 & 
[2\,1^2\,;2]
 \\
 (2^4\,1^2\,;1^2)
 & 
 48
 & 
\left( \begin{array}{@{}c@{}c@{}c@{}c@{}c@{}c@{}c@{}}
  0&&0&&1&&1 \\
 &1&&1&&2&
\end{array} \right)
 & 
[3\,1\,;1^2]
 \\
 (2\,1^4\,;2\,1^4)
 & 
 48
 & 
\left( \begin{array}{@{}c@{}c@{}c@{}c@{}c@{}}
  2&&2&&2 \\
 &0&&0&
\end{array} \right)
 & 
[2^2\,;2]
\\
 (2^3\,1^2\,;1^4)
 & 
 50
 & 
\left( \begin{array}{@{}c@{}c@{}c@{}c@{}c@{}}
 1&&1&&1 \\
 &1&&2&
\end{array} \right)
 & 
[3\,;1^3]
 \\
 (2^3\,1^2\,;2\,1^2)
 & 
 54
 & 
\left( \begin{array}{@{}c@{}c@{}c@{}c@{}c@{}}
 1&&2&&2 \\
 &0&&1&
\end{array} \right)
 & 
[3\,1\,;2]
 \\
 (3^2\,2\,1^2\,;1^2)
 & 
 54
 & 
\left( \begin{array}{@{}c@{}c@{}c@{}c@{}c@{}}
  1&&1&&1 \\
 &0&&3&
\end{array} \right)
 & 
[4\,1\,;1]
\end{array}
} \ee

\newpage

\be {\small
\begin{array}{l@{\hspace{50pt}}l@{\hspace{50pt}}c@{\hspace{50pt}}l}
(1^{13};\emp)
 & 
 0
 & 
\left(\ba{@{}c@{}c@{}c@{}c@{}c@{}c@{}c@{}c@{}c@{}c@{}c@{}c@{}c@{}c@{}}
 0&&0&&0&&0&&0&&0&&0 \\
 &1&&1&&1&&1&&1&&1&
\ea\right)
 & 
[1^6;\emp]
 \\[-0.5pt]
(1;1^{12})
 & 
 12
 & 
\left(\ba{@{}c@{}c@{}c@{}@{}c@{}c@{}c@{}c@{}c@{}c@{}c@{}c@{}}
 1&&1&&1&&1&&1&&1 \\
 &0&&0&&0&&0&&0&
\ea\right)
 & 
[1^5;1]
 \\[-0.5pt]
(2^2\,1^9;\emp)
 & 
 20
 & 
\left(\ba{@{}c@{}c@{}c@{}c@{}c@{}c@{}c@{}c@{}c@{}c@{}c@{}}
 0&&0&&0&&0&&0&&0 \\
 &1&&1&&1&&1&&2&
\ea\right)
 & 
[2\,1^4;\emp]
 \\[-0.5pt]
(1;2^2\,1^8)
 & 
 30
 & 
\left(\ba{@{}c@{}c@{}c@{}@{}c@{}c@{}c@{}c@{}c@{}c@{}}
 1&&1&&1&&1&&1 \\
 &0&&0&&0&&1&
\ea\right)
 & 
[1^3;1^3]
\\[-0.5pt]
(1^3;1^{10})
 & 
 30
 & 
\left(\ba{@{}c@{}c@{}c@{}@{}c@{}c@{}c@{}c@{}c@{}c@{}c@{}c@{}}
 1&&1&&1&&1&&1 \\
 &0&&0&&0&&1&
\ea\right)
 & 
[1^3;1^3]
 \\[-0.5pt]
(2^4\,1^5;\emp)
 & 
 32
 & 
\left(\begin{array}{@{}c@{}c@{}c@{}c@{}c@{}c@{}c@{}c@{}c@{}}
 0&&0&&0&&0&&0 \\
 &1&&1&&2&&2&
\end{array} \right)
 & 
[2^2\,1^2;\emp]
 \\[-0.5pt]
(3\,2^2\,1^6;\emp)
 & 
 36
 & 
\left(\begin{array}{@{}c@{}c@{}c@{}c@{}c@{}c@{}c@{}c@{}c@{}}
 0&&0&&0&&1&&1 \\
 &1&&1&&1&&1&
\end{array}\right)
 & 
[1^2;1^4]
 \\[-0.5pt]
(1^9,1^4)
 & 
 36
 & 
\left(\ba{@{}c@{}c@{}c@{}c@{}c@{}c@{}c@{}c@{}c@{}}
 0&&0&&0&&1&&1 \\
 &1&&1&&1&&1&
\ea\right)
 & 
[1^2;1^4]
 \\[-0.5pt]
(2^6\,1;\emp) 
 & 
 36
 & 
\left( \begin{array}{@{}c@{}c@{}c@{}c@{}c@{}c@{}c@{}}
 0&&0&&0&&0 \\
 &2&&2&&2&
\end{array} \right)
 & [2^3;\emp] 
 \\[-0.5pt]
(1;2^4\,1^4)
 & 
 40
 & 
\left(\ba{@{}c@{}c@{}c@{}@{}c@{}c@{}c@{}c@{}}
 1&&1&&1&&1 \\
 &0&&1&&1&
\ea\right)
 & 
[1;1^5]
 \\[-0.5pt]
(1^5;1^8)
 & 
 40
 & 
\left(\ba{@{}c@{}c@{}c@{}@{}c@{}c@{}c@{}c@{}c@{}c@{}}
 1&&1&&1&&1 \\
 &0&&1&&1&
\ea\right)
 & 
[1;1^5]
 \\[-0.5pt]
(1^7;1^6)
 & 
 42
 & 
\left(\ba{@{}c@{}c@{}c@{}c@{}c@{}c@{}c@{}}
 0&&1&&1&&1 \\
 &1&&1&&1&
\ea\right)
 & 
[\emp;1^6]
 \\[-0.5pt]
(1^3;2\,1^7)
 & 
 44
 & 
\left(\ba{@{}c@{}c@{}c@{}@{}c@{}c@{}c@{}c@{}c@{}c@{}}
 1&&1&&1&&1 \\
 &0&&0&&2&
\ea\right)
 & 
[3\,1^2;1]
 \\
(2^2\,1;1^8)
 & 
 44
 & 
\left(\ba{@{}c@{}c@{}c@{}@{}c@{}c@{}c@{}c@{}c@{}c@{}}
 1&&1&&1&&1 \\
 &0&&0&&2&
\ea\right)
 & 
[3\,1^2;1]
 \\[-0.5pt]
(1;3\,2^2\,1^5)
 & 
 44
 & 
\left(\ba{@{}c@{}c@{}c@{}@{}c@{}c@{}c@{}c@{}}
 1&&1&&2&&2 \\
 &0&&0&&0&
\ea\right)
 & 
[2\,1^2;2]
 \\[-0.5pt]
(2^21^5;1^4)
 & 
 48
 & 
\left(\ba{@{}c@{}c@{}c@{}c@{}c@{}c@{}c@{}}
 0&&0&&1&&1 \\
 &1&&1&&2&
\ea\right)
 & 
[3\,1;1^2]
 \\[-0.5pt]
(1;3\,2^4\,1)
 & 
 48
 & 
\left(\ba{@{}c@{}c@{}c@{}@{}c@{}c@{}}
 2&&2&&2 \\
 &0&&0&
\ea\right)
 & 
[2^2;2]
 \\[-0.5pt]
(2^21^3;1^6)
 & 
 50
 &
\left(\ba{@{}c@{}c@{}c@{}c@{}c@{}}
  1&&1&&1 \\
 &1&&2&
\ea\right)
 & 
[3;1^3]
 \\[-0.5pt]
(1^5;2^2\,1^4)
 & 
 50
 & 
\left(\ba{@{}c@{}c@{}c@{}@{}c@{}c@{}c@{}c@{}}
 1&&1&&1 \\
 &1&&2&
\ea\right)
 & 
[3;1^3]
 \\[-0.5pt]
(2^4\,1;1^4)
 & 
 52
 & 
\left(\ba{@{}c@{}c@{}c@{}c@{}c@{}}
 0&&1&&1 \\
 &2&&2&
\ea\right)
 & 
[3^2;\emp]
 \\[-0.5pt]
(1^3;3\,2^2\,1^3)
 & 
 54
 & 
\left(\ba{@{}c@{}c@{}c@{}@{}c@{}c@{}c@{}c@{}}
  1&&2&&2 \\
  &0&&1&
\ea\right)
 & 
[3\,1;2]
\\[-0.5pt]
(2^2\,1;2^2 \,1^4)
 & 
 54
 & 
\left(\ba{@{}c@{}c@{}c@{}@{}c@{}c@{}c@{}c@{}}
 1&&1&&1 \\
 &0&&3&
\ea\right)
 & 
[4\,1;1]
 \\[-0.5pt]
(1^5;3\,2^2\,1)
 & 
 56
 &  
\left(\ba{@{}c@{}c@{}c@{}@{}c@{}c@{}c@{}c@{}}
 0&&2&&2 \\
 &1&&1&
\ea\right)
 & 
[3;2\,1]
 \\[-0.5pt]
(2^2\,1;3\,2^2\,1)
 & 
 60
 & 
\left(\ba{@{}c@{}c@{}c@{}}
 2&&2 \\
 &2&
\ea\right)
 & 
[\emp;2^3]
\end{array}
}\non \ee

\begingroup\raggedright\endgroup

\end{document}